\documentclass[epj]{svjour}
\usepackage[]{graphicx}

\begin{document}
\title{A Method to Calculate Fission-Fragment Yields $Y(Z,N)$
versus Proton and Neutron Number
in the  Brownian Shape-Motion Model}
\subtitle{Application to calculations of U and Pu charge yields}
\author{Peter M\"{o}ller\inst{1} \and Takatoshi Ichikawa\inst{2}
\thanks{\emph{Present address:} Insert the address here if needed}%
}                     
\offprints{}          
\institute{Theoretical Division, Los Alamos National Laboratory,
Los Alamos, NM 87545, USA \and
Yukawa Institute for Theoretical Physics, Kyoto University,
Kyoto 606-8502, Japan}
\date{Received: date / Revised version: date}
%
\abstract{We propose a method to calculate the two-dimensional (2D)
  fission-fragment yield $Y(Z,N)$ versus both proton and neutron
  number, with inclusion of odd-even staggering effects in both
  variables. The approach is to use Brownian shape-motion on a
  macroscopic-microscopic potential-energy surface which, for a
  particular compound system is calculated
  versus four shape variables: elongation
  (quadrupole moment $Q_2$), neck $d$, left nascent fragment
  spheroidal deformation $\epsilon_{\rm f1}$, right nascent
  fragment deformation $\epsilon_{\rm f2}$ and two asymmetry
  variables, namely proton and neutron
  numbers in each of the two fragments. The extension of previous
  models  1) introduces a method to  calculate  this generalized potential-energy
  function and 2)  allows the correlated transfer of nucleon
pairs in one step, in addition to sequential transfer.
In the previous version  the potential energy was calculated as a function of $Z$
  and $N$ of the compound system and its shape, including
  the asymmetry of the shape.  We outline here how to
  generalize the model  from the
  ``compound-system'' model to a model where the emerging fragment
  proton and neutron numbers also enter, over and above the compound
  system composition.
\PACS{
      {25.85.Ca}{discribing text of that key}   \and
      {24.10.Lx}{discribing text of that key}   \and
      {24.75.+i}{discribing text of that key}   \and
      {25.85.Jg}{discribing text of that key}
     } 
} 
\authorrunning{P. M\"oller and T. Ichikawa}
\titlerunning{A Method to Calculate Fission-fragment Yields $Y(Z,N)$}
\maketitle
\section{Introduction}
\label{intro}
 
In previous investigations it has been shown that a realistic
description of the experimentally observed fission-fragment charge
distributions can be obtained by means of random walks on tabulated
five-dimensional (5D) potential-energy surfaces calculated for a
densely spaced grid for over five million different shapes
\cite{randrup11:a,randrup11:b,randrup13:a} .  It was particularly
encouraging that the 70 charge-yield distributions measured at GSI
\cite{schmidt00:a} were well reproduced, including the transition from
symmetric fission for light Th isotopes to asymmetric fission for the
heavier isotopes beyond $A \approx 226$.  However, the often strongly
fluctuating patterns of odd-even staggering were not possible to
obtain in that implementation of the model. Briefly stated, the reason
was that there was no mechanism that allowed the properties of the two
nascent fragments to affect the calculated potential-energy
surface. For example, pairing was treated for the compound system as a
whole.
 
In the literature there have been numerous discussions of odd-even
staggering and models proposed to quantitatively describe this
feature. For example the effect has been correlated with
Coulomb-related quantities, such as $Z^2/A$ and $Z^2/A^{1/3}$,
referred to as order parameters, and to pairing effects on the nuclear
level density and its variation with excitation energy,
see Refs.~\cite{bocquet89:a,rejmund00:a,caamano11:a} and references
therein. Other  models are based on properties of separated
fragments and thermal equilibrium at scission. For a review see
Ref.~\cite{benlliure98:a}. Common for these models are that they are
not based on detailed, calculated potential-energy surfaces or
dynamical evolution on such surfaces. The models often also contain a
substantial number of postulated terms with parameters which are
determined from adjustments to observed yields.
Another group of  models do treat dynamical evolution,
in a Langevin approach. Until now they are based on
macroscopic potential-energy surfaces. or, when shell effects
are included the calculations are performed for
three shape variables for fairly high excitation
energies ($E^* = 20$ MeV), see Ref.\ \cite{moller15:b} for a brief review
but with extensive references to original work.
 
In the Brownian shape-motion (BSM) model the only
parameters are those of a well-established macro\-scopic-micro\-sco\-pic
model used to calculate the five-dimensional (5D) potential-energy
surfaces \cite{moller09:a} in which the parameters have been unchanged
since 2002 \cite{moller04:a}, the additional critical neck radius
at which the mass split is frozen, a weak bias potential
(but the results are fairly
insensitive both to the magnitude of the critical neck radius
and to the bias potential strength \cite{randrup11:a,randrup11:b}),
and two parameters in a ``suppression factor'' that accounts
for the decrease of the shell-plus-pairing correction with
energy \cite{randrup13:a}.
 
We have
recently shown that the observed magnitude of the odd-even staggering
can be  directly correlated to the excitation energy above the outer part
of the calculated potential-energy surface \cite{moller14:b}.
Therefore we suggested the BSM model
could describe odd-even staggering if a potential-energy model were
developed that accounts for how the individual nascent fragment
properties are expressed in the calculated  potential-energy surface
\cite{moller14:b}.
 
Here we propose a model for the potential-energy surface in which
the properties of the individual fragments gradually emerge as
the scission configuration is approached and specify the full
details this proposed model. To treat odd-even staggering
we add to the potential energy, as is
customarily done since the dawn of  nuclear mass calculations
\cite{weizsacker35:a,bethe36:a}, a pairing contribution $\Delta$
to emerging odd fragments.
However, since we start our
trajectories at the ground-state shape of the even-even parent nuclei,
where there is no odd pairing effect, only a fraction of the  pairing
delta of the fully formed final fragments is added in the initial stages
of division. This fraction grows with decreasing neck diameter, see below for
further discussion and specification.
 
We also introduce a correlated transfer of paired nucleon configurations.
Such correlated nucleon transfers, which are different from sequential
transfers, are often identified in nuclear reaction experiments,
for a discussion see the presentation in Ref.\ \cite{broglia12:a}
and the many references cited therein, for example Refs.\ \cite{oertzen01:a,corradi11:a}.
 
It turns out that we can incorporate our new ideas in calculations of charge yields, with
minimal modifications of existing computer codes. Our first results on charge distributions
are presented below.
 
\section{Shape parameterization and notations}
 
To describe the nuclear shape we use the three-quadratic-surface (3QS)
parameterization. It was introduced almost 50 years ago \cite{nix69:a}.
It  is much more cumbersome to deal with than, say,
a multipole expansion such as the $\beta$ parameterization
\cite{moller95:b}, but is used to allow a realistic description of
shapes of a fissioning nucleus up to and including division of
the single shape into separated fragments.
Particularly noteworthy is that the emerging fragment shapes can
be deformed spheroids or exact spheres. The latter is of special
importance because it allows the extra binding associated
with {\it spherical} doubly-magic nuclei to be accurately calculated.
More details are  found in Ref.\ \cite{nix69:a}; the discussions
of Figs.\ 1 and 2 there are particularly informative. How we design our potential-energy
calculations in terms of this parameterization is detailed in Refs.\ \cite{moller01:a,moller09:a};
here we briefly summarize a few essential details.
 
In the (3QS) parameterization the
shape of the nuclear surface is specified in terms of three
smoothly joined portions of quadratic surfaces
of revolution. They are completely specified \cite{nix69:a} by
\begin{eqnarray}
\rho(z)^2= \left\{ \begin{array}{ll}
{a_1}^2-
\begin{displaystyle}\frac{{a_1}^2}{{c_1}^2}\end{displaystyle}
(z-l_1)^2 \; \; ,&l_1-c_1\leq z \leq z_1\\[2ex]
{a_2}^2-
\begin{displaystyle}\frac{{a_2}^2}{{c_2}^2}\end{displaystyle}
(z-l_2)^2 \; \; ,&z_2 \leq z \leq l_2+c_2 \\[2ex]
{a_3}^2-
\begin{displaystyle}\frac{{a_3}^2}{{c_3}^2}\end{displaystyle}
(z-l_3)^2 \; \; ,& z_1 \leq z \leq z_2
\end{array}
\right.
\label{threeqs}
\end{eqnarray}
Here the left-hand surface is denoted by the subscript 1, the
right-hand one by 2 and the middle one by 3.  Shapes 1 and 2 are
spheroids, for which $c$ is the semi-symmetry axis length, $a$ is
the semi-transverse axis length, and $l$ specifies the location of the
center of the spheroid.  The middle body may be a spheroid or a hyperboloid of
one sheet, for which $c_3$ is imaginary.  At the left and right
intersections of the middle surface with the end surfaces the value
of $z$ is $z_1$ and $z_2$, respectively.  Surfaces 1 and 2 are also
referred to as end bodies and, alternatively, nascent fragments.
\begin{figure}[t]
 \begin{center}
\includegraphics[width=\linewidth]{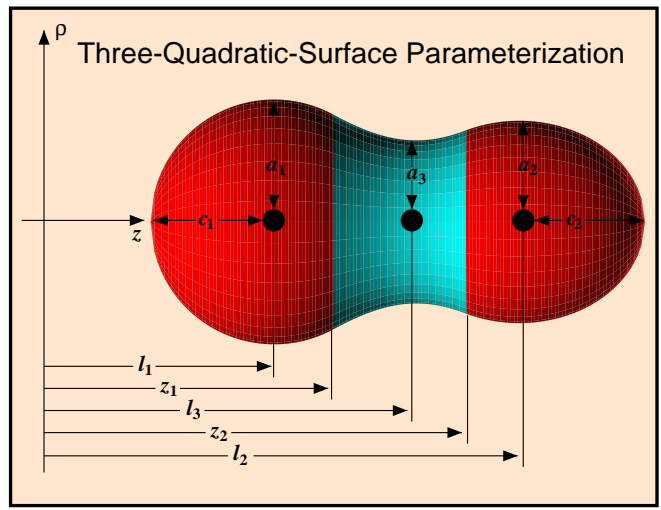}
\caption{Shape generated by the three-quadratic-surface parameterization.
Different colors distinguish between the shape sections generated by the three
expressions in \vspace{-0.3in}Eq.\ (\ref{threeqs}). }
\label{threeqsshape}
 \end{center}
\end{figure} 	
A shape generated by the parameterization in Eq.\ (\ref{threeqs})
is shown in Fig.\ \ref{threeqsshape}.
 
In our calculations we use five shape parameters:
elongation in terms of quadrupole moment $Q_2$,
left and right fragment spheroidal deformations $\epsilon_{\rm f1}$
and $\epsilon_{\rm f2}$, neck diameter $d$ and mass asymmetry $\alpha_{\rm g}$.
These five parameters completely specify the shape for which
the potential energy is calculated and have been extensively discussed
in Ref.\ \cite{moller09:a}. Also, they completely exhausted the shape space
available to this parameterization (except, trivially, for including larger values in each
of the five shape parameters). For our  studies here
we need to revisit the asymmetry variable $\alpha_{\rm g}$
which is directly connected to the asymmetry of nuclear shape
(and the separated fragments):
\begin{equation}
\alpha_{\rm g} = \frac{a_1^2 c_1 - a_2^2 c_2}
                      {a_1^2 c_1 + a_2^2 c_2}.
\label{asymdefg}
\end{equation}
This is equivalent to
\begin{equation}
\alpha_{\rm g} = \frac{M_1^{\rm c} - M_2^{\rm c}}{M_1^{\rm c}+M_2^{\rm c}}
\label{asymdef}
\end{equation}
where $M_1^{\rm c}$ and $M_2^{\rm c}$ are the  volumes inside the
end-body quadratic surfaces, were they
completed to form closed-surface spheroids and ``c'' in the notation
is to clarify we are referring to quantities of the compound single-shape system.
To avoid the introduction of a large number of equivalent concepts we
use somewhat interchangeably, ``mass'', ``volume'' and nucleon number $A$.
 
Our random-walk tracks end at a point where the neck radius of the nuclear
shape is quite well developed, namely at 2.5 fm (compare to the radius
of a spherical $^{240}$Pu nucleus which is 7.2 fm), so the shape has almost
reached the configuration of separated fragments. The neck radius is
actually smaller than the radius of $^{16}$O for which the radius is 2.92 fm,
and which is the lightest nucleus to which we have applied the macroscopic-microscopic method.
Although $M_1^{\rm c}$ and $M_2^{\rm c}$ for shapes with well-developed
necks can be expected to be close to the final fragment
masses
we cannot directly compare $M_1^{\rm c}$ and $M_2^{\rm c}$ to
the observed fission fragment masses
$M^{\rm s}_1$ and $M^{\rm s}_2$ (where the superscript ``s'' indicates
we refer to the separated fragments) because the former do not quite
sum up to the total nuclear volume or mass $A$. However, by scaling
$M_1^{\rm c}$ and $M_2^{\rm c}$
so that their sum  adds up to the total
mass number $A$, we can directly relate  the ``mass asymmetry''
of the  compound-nucleus shape to the observed heavy and light fragment masses.
We obtain trivially
\begin{eqnarray}
M_1^{\rm s} &=& r_{\rm s}M_1^{\rm c}=A\frac{1+\alpha_{\rm_g}}{2}
\nonumber \\
M_2^{\rm s} &=& r_{\rm s}M_2^{\rm c}=A\frac{1-\alpha_{\rm_g}}{2}
\; \; \;
{\rm and} \\
r_{\rm s} &=& \frac{A}{M_1^{\rm c} + M_2^{\rm c}}
\nonumber
\label{asymsings}
\end{eqnarray}
where $r_{\rm s}$ is a scaling factor
for a nucleus with $A$ nucleons. The scaling
is equivalent to a assigning
the mass in the neck region to the left and right nascent fragments in
proportion to the respective volumes of these nascent fragments,
were they completed to closed
spheroids. The amount of matter involved is in the range of
10--20 nucleons.
It is obvious how the same definitions apply to the proton and neutron numbers.
The symbols  $Z$ and $N$ are used for
proton and neutron numbers of the fissioning compound system.
We then have:
\begin{equation}
Z_2^{\rm s} = Z - Z_1 ^{\rm s}\; \; \; \; \; {\rm and} \; \; \; \;  N_2^{\rm s} = N -N_1^{\rm s}
\end{equation}
Therefore we usually use only $Z_1^{\rm s}$ and $N_1^{\rm s}$ when we
explicitly refer to the number of fragment
protons and neutrons, the other fragment proton and neutron numbers are
then also specified.
To calculate the yield as a function of both fragment proton and neutron number  obviously
requires that  the previous 5D model (which always assumed that the proton to neutron ratios in
both fragments were equal to the proton to neutron ratio in the compound system) be generalized
to 6D so that  the yield $Y(Z_1^{\rm s},N_1^{\rm s},Z_2^{\rm s},N_2^{\rm s})$, which is a
function of both proton and neutron asymmetry is obtained.
We also realize that to describe odd-even effects requires that we space the
grids we will use in terms of integer spacing of $Z_1^{\rm s}$ and $N_1^{\rm s}$ in some fashion which
we will now introduce.
 
\section{Potential energy versus shape and nascent fragment proton and neutron number}
The method  to calculate the two-dimensional fission-fragment yield $Y(Z,N)$
function that we now introduce is perhaps most transparently explained by
occasionally referring to specific aspects of the computer code used to calculate,
in the macroscopic-microscopic method, potential-energies as  functions
of shape.   In our current
potential-energy  model and code  we obtain
a total potential energy for a specific compound nucleus $(Z,N)$
and a specific   ``shape'' as a sum of
a macroscopic energy (given by a liquid-drop type expression)
and a microscopic shell-plus-pairing correction:
\begin{eqnarray}
E_{\rm pot}(Z,N,{\rm shape}) = E_{\rm mac}(Z,N,{\rm shape}) \nonumber\\[1ex]
 +E_{\rm s+p}^{\;{\rm prot}}(Z,(N),{\rm shape}) \\[1ex]
+ E_{\rm s+p}^{\;{\rm neut}}((Z),N,{\rm shape})\nonumber
\end{eqnarray}
where $ E_{\rm mac}(Z,N,{\rm shape})$ is the macroscopic energy
and $E_{\rm s+p}^{\;{\rm prot}}$ is the proton shell-plus-pairing
correction and the final term the neutron shell-plus-pairing
correction. In a calculation for a specific compound system
and a specific shape these shell-plus-pairing corrections
 can trivially be
individually tabulated. To obtain, say, the proton microscopic correction
single-particle levels are calculated numerically in a folded-Yukawa
single-particle potential with a functional form derived from  the nuclear shape
\cite{bolsterli72:a,moller95:b}. From these levels the shell correction is obtained
by use of the Strutinsky method \cite{strutinsky67:a,strutinsky68:a} and
the pairing correction through, in our case, the Lipkin-Nogami method,
as detailed in \cite{moller92:c}. Thus, the proton microscopic correction
is independent of neutron number, except that the potential radius and depth
depend on both proton and neutron number, therefore we have used
the notation $(N)$ and $(Z)$ to indicate a weak dependence.  But for small variations of
neutron number the effect on the proton {\it microscopic} correction is
small and can be neglected, an important observation that we will make use
of later. To show this insensitivity we have calculated the proton and neutron
shell-plus-pairing corrections for the ground-ground state shape of $^{270}_{108}$Hs$_{162}$ for
the single-particle fields appropriate for this nucleus. We find for the
proton and neutron  shell-plus-pairing corrections -3.7023 and -5.3715 MeV, respectively.
When we do the calculation for single-particle fields appropriate for $^{274}_{112}$Cn$_{162}$ we
find that the proton and neutron shell-plus-pairing corrections are -3.6305 and -5.3353 MeV
respectively. So when we change the proton number by 4 units and implement
the corresponding effect on the neutron single-particle field
the neutron shell correction changes
by only 0.0362 MeV. The change in the proton shell-plus-pairing correction is larger
because proton field changes   more than the neutron field
when we change the proton
number.
 
In the fission potential-energy code the shapes that can be studied
are described in terms of the three quadratic surfaces of revolution: two
end spheroidal sections and a middle region that near scission is a
hyperboloid of revolution \cite{bolsterli72:a} as discussed in
the previous section, see Eq.\ (\ref{threeqs}) and Fig.\ \ref{threeqsshape}.
In the computer code
the equations for these shapes need to be specified; five independent
shape parameters are required for the shape specification.
Historically the shape parameters $\alpha$ and $\sigma$ were used
\cite{bolsterli72:a}. But to more
directly visualize the shape, we now characterize the shape in terms
of  five equivalent shape parameters: quadrupole moment
$Q_2$, related to the elongation of the shape, neck diameter $d$, left
fragment spheroidal deformation $\epsilon_{\rm f1}$, right fragment
spheroidal deformation $\epsilon_{\rm f2}$, and mass asymmetry
$\alpha_{\rm g}$. The transformations from these parameters to  the
parameters of the quadratic functions that generate the shapes in the
code are very lengthy and  non-linear. They are described in
Ref.\ \cite{moller09:a}.  For our discussion here we need to know that
\begin{equation}
\alpha_{\rm g} = \frac{M_1^{\rm c} - M_2^{\rm c}}{M_1^{\rm c}+M_2^{\rm c}}
=\frac{M_1^{\rm s} - M_2^{\rm s}}{M_1^{\rm s}+M_2^{\rm s}}
\label{asymdef2}
\end{equation}
 
Rather than discussing the asymmetry in terms of nucleon
number $A$ we can use the above concepts and
discuss the charge asymmetry; we have earlier assumed
that the proton to neutron ratio is the same in both
fragments. We now develop an approach to
treat different ratios, that is if the proton and neutron
numbers are $Z_1^{\rm s}$ and $N_1^{\rm s}$ in one of  the fragments and
$Z-Z_1^{\rm s}$ and $N-N_1^{\rm s}$ in the other fragment we will treat
\begin{equation}
\frac{Z_1^{\rm s}}{N_1^{\rm s}} \neq \frac{Z}{N} \neq \frac{Z-Z_1^{\rm s}}{N-N_1^{\rm s}}
=\frac{Z_2^{\rm s}}{N_2^{\rm s}}
\end{equation}
It follows from previously that  when we discuss the asymmetry in terms
of proton number we can write
\begin{equation}
\alpha_{\rm g} = \frac{Z_1^{\rm s} - Z_2^{\rm s}}{Z_1^{\rm s}+Z_2^{\rm s}}
\label{asymdefcharge}
\end{equation}
When we calculate the potential energy for the compound system
that should correspond to specific (in our case integer) separated-fragment charge numbers
we use Eq.\ (\ref{asymdefcharge}) to define the asymmetry $\alpha_{\rm g}$
of the corresponding compound-nucleus shape for which we calculate
proton shell-plus-pairing corrections  and  macroscopic energies
which are needed in our model. We need additional terms
to describe how the macroscopic energy changes when
we allow different proton to neutron ratios in
the two fragments, mainly a symmetry-energy effect. We will discuss
how to obtain this effect below. Correspondingly we can define the
asymmetry $\alpha_{\rm g}$  for neutrons so that it corresponds
to integer splits of neutron number and tabulate the calculated
neutron shell-plus-pairing corrections. Below we specify how these
results serve as the starting point to obtain the potential energy
for ratios between the proton and neutron numbers that are different in the two fragments.

We discussed above why the shell-plus-pairing corrections for protons and
neutrons can be calculated independently of each other, to a very
high degree of accuracy. Therefore  we can
write
\begin{eqnarray}
\lefteqn{E_{\rm pot}(Z,N,Q_2,d,\epsilon_{\rm f1},\epsilon_{\rm f2},Z_1^{\rm s},N_1^{\rm s}) =} \nonumber \\[1ex]
& &  E_{\rm mac}(Z,N,Q_2,d,\epsilon_{\rm f1},\epsilon_{\rm f2},Z_1^{\rm s},N_1^{\rm s}) \nonumber\\[1ex]
 &+&E_{\rm s+p}^{\;{\rm prot}}(Z,N,Q_2,d,\epsilon_{\rm f1},\epsilon_{\rm f2},\alpha_{\rm g}(Z_1^{\rm s}))  \\ [1ex]
&+& E_{\rm s+p}^{\;{\rm neut}}(Z,N,Q_2,d,\epsilon_{\rm f1},\epsilon_{\rm f2},\alpha_{\rm g}(N_1^{\rm s})) \nonumber \\[1ex]
&+& E_{\rm odd}\nonumber
\label{pottot}
\end{eqnarray}
 Therefore, to obtain the total shell-plus-pairing corrections
for any fragment split ($Z_1^{\rm s},N_1^{\rm s})$ we calculate and tabulate the proton shell-plus-pairing
corrections for a grid in $\alpha_{\rm g}$  corresponding to
integer $Z_1^{\rm s}$ (and the corresponding $Z_2^{\rm s}$), obtained from Eq.\ (\ref{asymdefcharge}).
We calculate the neutron shell-plus-pairing
correction for a different spacing in $\alpha_{\rm g}$ corresponding to
integer spacing in $N_1^{\rm s}$. Thus the 6-dimensional shell-plus-pairing correction for any mass
split $(Z_1^{\rm s},N_1^{\rm s})$ is the sum of two 5-dimensional functions.

To obtain
$E_{\rm mac}(Z,N,Q_2,d,\epsilon_{\rm f1},\epsilon_{\rm f2},Z_1^{\rm s},N_1^{\rm s})$
we proceed in several steps.
First, when
we run the code to calculate and tabulate the proton shell-plus-pairing correction
for integer values of $Z_1^{\rm s}$ we  also tabulate the macroscopic energy. It will
then be obtained  for  non-integer values
\begin{equation}
N_1^{\rm t} = N\times\frac{Z_1^{\rm s}}{Z}
\end{equation}
of $N_1^{\rm s}$,
because the asymmetry variable $\alpha_{\rm g}$ was chosen
to correspond to integer values of $Z_1^{\rm s}$.
 
Thus we have tabulated
\begin{equation}
 E_{\rm mac}(Z,N,Q_2,d,\epsilon_{\rm f1},\epsilon_{\rm f2},Z_1^{\rm s},N_1^{\rm t})
\label{compmacen}
\end{equation}
where we need to remember that here $N_1^{{\rm t}}$
corresponds to a non-integer value because the asymmetry of the shape
was chosen to yield integer $Z_1^{\rm s}$. The superscript ``t'' stands
for ``tabulated''. We need this tabulated value as one term  in the
macroscopic energy-model expression we now develop.
 
But we need to obtain the macroscopic energy for (several different) integer
$N_1^{{\rm s}}$.  It would be difficult to obtain such macroscopic energies
by developing a model that integrated across the compound nuclear shape
for a configuration with variable proton and neutron densities across
the shape. But we now pose that we will get a sufficiently accurate
model by considering {\it changes} in the macroscopic energy
relative to the tabulated macroscopic energy we discussed in Eq.\
(\ref{compmacen}). These changes are mainly due to changes in
the symmetry energies, with considerably smaller contributions
from other effects.
 We can obtain those by suitable
consideration of  changes in the macroscopic energy of separated fragments.
These we calculate as {\it changes} in sum of the macroscopic
energy of separated spherical fragments.
Therefore we calculate the sum of the {\it spherical} macroscopic energies for
two separated nuclei
for  the specific fixed $Z_1^{\rm s}$ (related to Eq.\ (\ref{compmacen})) and for a number of different
 $N_{\nu}$:
\begin{eqnarray}
\lefteqn{E_{\rm mac}^{\;{\rm sep}}(Z_1^{\rm s},N_{\nu},Z-Z_1^{\rm s},N-N_{\nu}) =} \nonumber\\[1ex]
 & & E_{\rm mac}^{\;{\rm sph}}(Z_1^{\rm s},N_{\nu}) + E_{\rm mac}^{\;{\rm sph}}(Z-Z_1^{\rm s},N-N_{\nu})
\label{fragmacsum}
 \end{eqnarray}
No odd-particle pairing effects should be included here; those
are treated as discussed below.
This function is tabulated for $Z_1^{\rm s} = 52$ in Table 1 for fission of
$^{236}$U.  We note that in this integer-spaced grid the minimum
energy occurs for a split where the sum of  $Z/N$ ratios in
the two fragments  is as close as possible to $2\times Z/N$ of
the compound nucleus. The line corresponding to this split is
indicated by a ``C'' at the very right.
\begin{table}
\begin{center}
 \caption{Spherical fragment macroscopic energies and
their sums for various fragmentations
of the compound system $^{236}$U leading to fragment charges 52/40;
only the neutron fragmentations vary. the columns labeled
$E_{\rm f1}$ and $E_{\rm f2}$ correspond the first and second
term of line two in Eq.\ (\ref{fragmacsum}).
The lowest sum is obtained with the
proton to neutron ratio as equal as possible in the two fragments and to that of the compound
system. The line corresponding to this division is indicated by
a ``C'' in the last column.}
\begin{tabular}{rrrrrrr}
\hline\\
   $Z_1^{\rm s}$& $N_1^{\rm s}$&
\multicolumn{1}{c}{$E_{\rm f1}$} & $Z_2^{\rm s}$& $N_2^{\rm s}$&
\multicolumn{1}{c}{$E_{\rm f2}$}&
\multicolumn{1}{c}{$E_{\rm f1}+E_{\rm f2}$}\\
     &   &       &     &   &         &\\
   52& 96& -15.95&   40& 48& -84.661 &  -100.608  \phantom{C}\\
   52& 94& -26.35&   40& 50& -87.387 &  -113.741  \phantom{C}\\
   52& 92& -36.10&   40& 52& -88.671 &  -124.770  \phantom{C}\\
   52& 90& -45.16&   40& 54& -88.592 &  -133.751  \phantom{C}\\
   52& 88& -53.51&   40& 56& -87.222 &  -140.731  \phantom{C}\\
   52& 86& -61.12&   40& 58& -84.626 &  -145.748  \phantom{C}\\
   52& 84& -67.97&   40& 60& -80.868 &  -148.839  \phantom{C}\\
     &   &       &     &   &         &            \phantom{C}\\
   52& 82& -74.03&   40& 62& -76.004 &  -150.032C\\
     &   &       &     &   &         &            \phantom{C}\\
   52& 80& -79.26&   40& 64& -70.088 &  -149.348  \phantom{C}\\
   52& 78& -83.64&   40& 66& -63.170 &  -146.807  \phantom{C}\\
   52& 76& -87.12&   40& 68& -55.298 &  -142.422  \phantom{C}\\
   52& 74& -89.68&   40& 70& -46.515 &  -136.199  \phantom{C}\\
   52& 72& -91.28&   40& 72& -36.862 &  -128.143  \phantom{C}\\
   52& 70& -91.87&   40& 74& -26.377 &  -118.250  \phantom{C}\\
   52& 68& -91.42&   40& 76& -15.097 &  -106.515  \phantom{C}\\
   52& 66& -89.87&   40& 78&  -3.055 &   -92.926  \phantom{C}\\
\hline\\[-5ex]
\end{tabular}
\label{summac}
\end{center}
\end{table}
\begin{figure}[t]
 \begin{center}
\includegraphics[width=\linewidth]{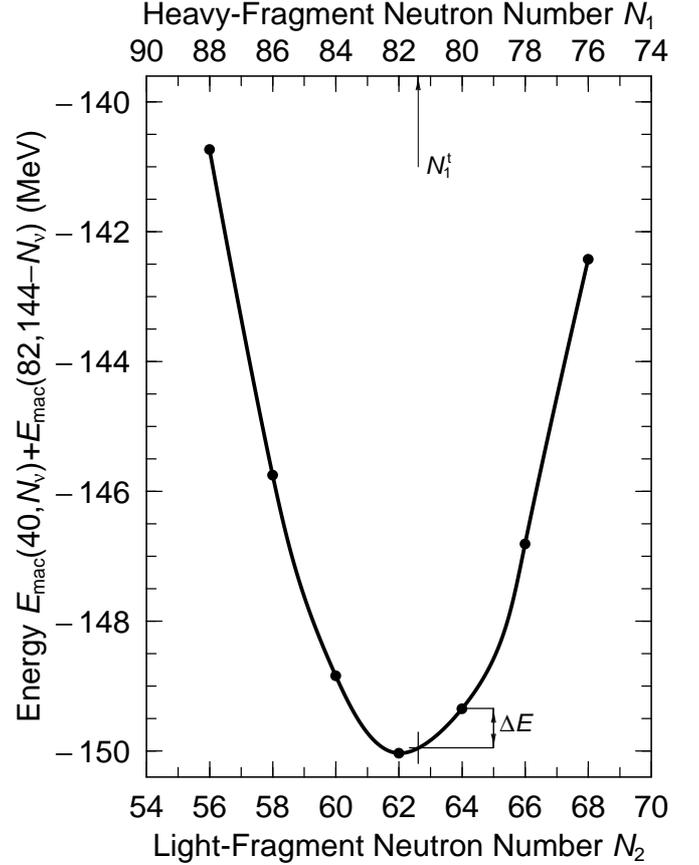}
\caption{Sum of separated-fragment macroscopic \vspace{-0.25in}energies.}
\label{easym}
 \end{center}
\end{figure} 	
 
We  pose that the macroscopic energy for any fragment division $(Z_1^{\rm s},N_1^{\rm s})$
in the fissioning system is given as
\begin{eqnarray}
\lefteqn{E_{\rm mac}(Z,N,Q_2,d,\epsilon_{\rm f1},\epsilon_{\rm f2},Z_1^{\rm s},N_1^{\rm s}) =} \nonumber \\[1ex]
& & E_{\rm mac}(Z,N,Q_2,d,\epsilon_{\rm f1},\epsilon_{\rm f2},Z_1^{\rm s},N_1^{\rm t})\nonumber \\[1ex]
&+ &  E_{\rm mac}^{\;{\rm sep}}(Z_1^{\rm s},N_{\nu},Z-Z_1^{\rm s},N-N_{\nu}) \nonumber \\[1ex]
&- & E_{\rm mac}^{\;{\rm sep}}(Z_1^{\rm s},N_{1}^{\;{\rm t}},Z-Z_1^{\rm s},N-N_1^{\;{\rm t}})
\label{mtot}
\end{eqnarray}
where the last term is calculated by interpolation in the table
corresponding to Eq.\ (\ref{fragmacsum}). As a specific example we discuss
a fragment division where the charge split is 52/40. Then $N_1^{\rm t}=81.39$.
We tabulate the sum in Eq.\ (\ref{fragmacsum}) as the right column in Table \ref{summac} and plot
this sum for the specific charge division in our example in Fig.\ \ref{easym}.
As an example that we can now, in our model, calculate the macroscopic energy
for any $(Z_1^{\rm s},N_1^{\rm s})$. To illustrate this  we continue to discuss our specific example.
Equation \ref{mtot} is a complete specification of the method. Suppose we want to calculate
the macroscopic energy for the particular case
$E_{\rm mac}(92,144,Q_2,d,\epsilon_{\rm f1},\epsilon_{\rm f2},52,64)$.
Then we obtain the first term in the right member of Eq.\
(\ref{mtot}) from our tabulated function in Eq.\ (\ref{compmacen}).
For our choice of $N_1^{\rm s}=64$ the second term is given
by the energy at the upper horizontal line and the third term by the energy
at the lower horizontal line in Fig.\ \ref{easym}. Thus in this
example we obtain the macroscopic energy as a sum of the tabulated
macroscopic energy plus $\Delta E$ indicated in Fig.\ \ref{easym}.
 
One may argue that since some of the fissioning shapes involve
deformed nascent fragments the terms in Eq.\ (\ref{fragmacsum}) should
be evaluated for the corresponding deformed shapes. But for the {\it differences}
we consider here it only makes a minuscule difference. Let us choose
$^{102}$Zr and $^{106}$Zr in Table \ref{summac} as an illustrative example.
The energy difference for the spherical shapes $\Delta E_{\rm sph}$  tabulated is
\begin{equation}
\Delta E_{\rm sph} =-63.108 -(-75.973)= 12.865
\end{equation}
We have evaluated the macroscopic energies for deformed
shapes with $\epsilon_2= 0.5$, which is the largest deformation for
emerging fragments in our specified deformation grid, see Ref.\ \cite{moller09:a} and obtain
\begin{equation}
\Delta E_{\rm def} =-53.262  -(-66.108)= 12.846
\end{equation}
Thus, although the absolute energies change considerably
the effect on their differences is completely negligible,
only 0.019 MeV\@.
 
Finally we need to account
for odd pairing effects in the emerging fragments.
When the fragments are fully developed with zero neck radius of the
compound system a common assumption is that for odd-odd splits the
extra odd contribution to the  energy should be
$E_{\rm odd} = 2\times \Delta$
where $\Delta$ is the pairing-gap parameter, chosen as $\Delta=1.0$ MeV\@.
With a non-zero neck radius the effect of pairing of the emerging odd
splits would be smaller; for very compact shapes with no
obvious neck there should be no effect. This leads to the following
prescription for the odd energy of the fissioning system
\begin{eqnarray}
\begin{array}{lclr}
E_{\rm odd} & = &  2\times \Delta \times (B_{\rm W}-1)^k & {\rm  \;odd}\; \;Z_1^{\rm s},Z_2^{\rm s}\\[1ex]
E_{\rm odd} & = & 0 & {\rm  \;even}\; \;Z_1^{\rm s},Z_2^{\rm s}\\
\end{array}
\label{pair}
\end{eqnarray}
where, with the choice $k=1$, $(B_{\rm W} -1)^k$ is the shape dependence of the Wigner term in our
potential-energy model.  The shape factor $B_{\rm W}$ is 1
for a shapes with no neck and increases continuously, as the neck
develops, to 2 for separated fragments. It is necessary to postulate
such a shape dependence because the macroscopic energy of two separated
fragments contains two Wigner terms, the original system only one.
Without such a shape dependence a discontinuity of the order of 10 MeV would
occur at scission of actinide nuclei. That is, if we calculated the energy for
a deforming nucleus up to the scission point we would at scission
obtain a 10 MeV lower energy than if we calculated the energy
of two approaching separated fragments.
A pedagogical figure illustrating this and the necessity of this shape dependence
is in Ref.~\cite{moller04:a}, Fig.~1.
Since we need a realistic potential-energy surface in the scission region
we do need to consider these issues (which have in many investigations been ignored).
The comprehensive discussion of the shape-dependence
of the Wigner term in Ref.~\cite{moller89:a} carries directly
over to how the effect of the pairing $\Delta$ increases as the neck
becomes more narrow. There is no known derivation of the Wigner shape dependence
so it is just postulated, but with consideration of its limiting behavior \cite{moller89:a}.
The power constant $k$, which we introduced here to allow some
sensitivity studies, governs how early in the division process
the character of the two fragments  causes the ``second'' Wigner term,
or in our case, the odd pairing effect, to manifest itself.
Below we will present  sensitivity studies on the shape dependence
and on the magnitude of the pairing delta.

\begin{figure}[t]
 \begin{center}
\includegraphics[width=0.65\linewidth]{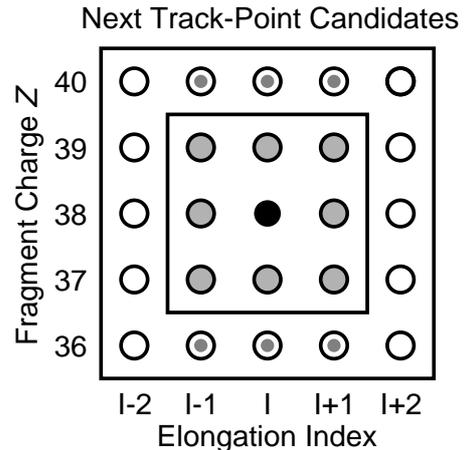}
   \caption{Two-dimensional slice in the 5D space schematically illustrating
possible candidate points for the next trajectory step.
The smaller square with gray-filled circles  inside, indicates
the limit originally chosen  for next step candidate points.
The circles partially filled
with gray correspond to  transfer of paired configurations.
\vspace{-0.2in}}
\label{grid}
 \end{center}
\end{figure} 	
\begin{figure}[t]
 \begin{center}
\includegraphics[width=\linewidth]{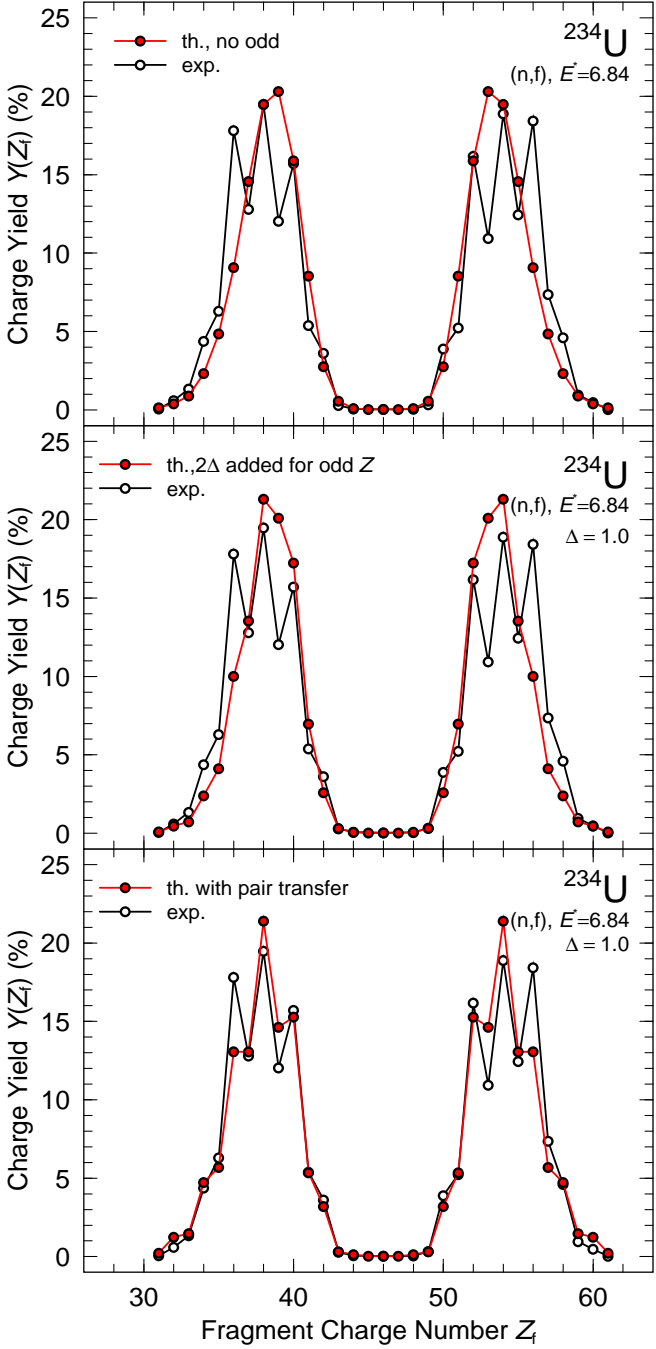}
\caption{Calculated and measured charge yields for neutron-induced fission of
$^{234}$U. The  top panel represents the original model,
the middle panel has $2\times \Delta$ added to the potential
energy in the matrix locations corresponding to odd charge splits.
In the bottom panel steps corresponding to two-proton changes
in asymmetry \vspace{-0.3in} are allowed.}
\label{yields1}
 \end{center}
\end{figure} 	
\begin{figure}[t]
 \begin{center}
\includegraphics[width=\linewidth]{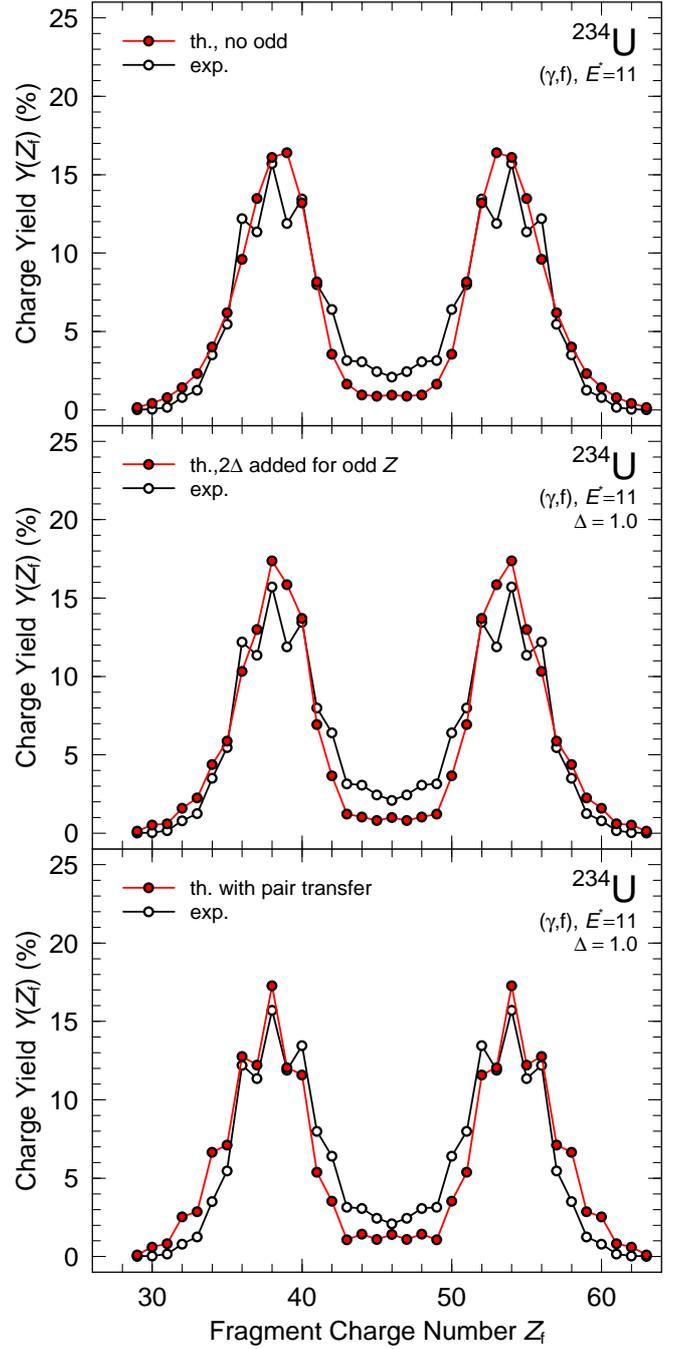}
\caption{As  Fig.~\ref{yields1}, but the experimental data
are for ($\gamma$,f) reactions
leading to $E^*\approx 8-14$~MeV;
they include contamination from fission after
1n($\approx15\%$) and 2n($\approx5\%$) emission\cite{schmidt00:a}.
The calculations are for fission of $^{234}$U at $E^* =11$ MeV.
\vspace{-0.3in}}
\label{yields2}
 \end{center}
\end{figure} 	
\begin{figure}[t]
 \begin{center}
\includegraphics[width=\linewidth]{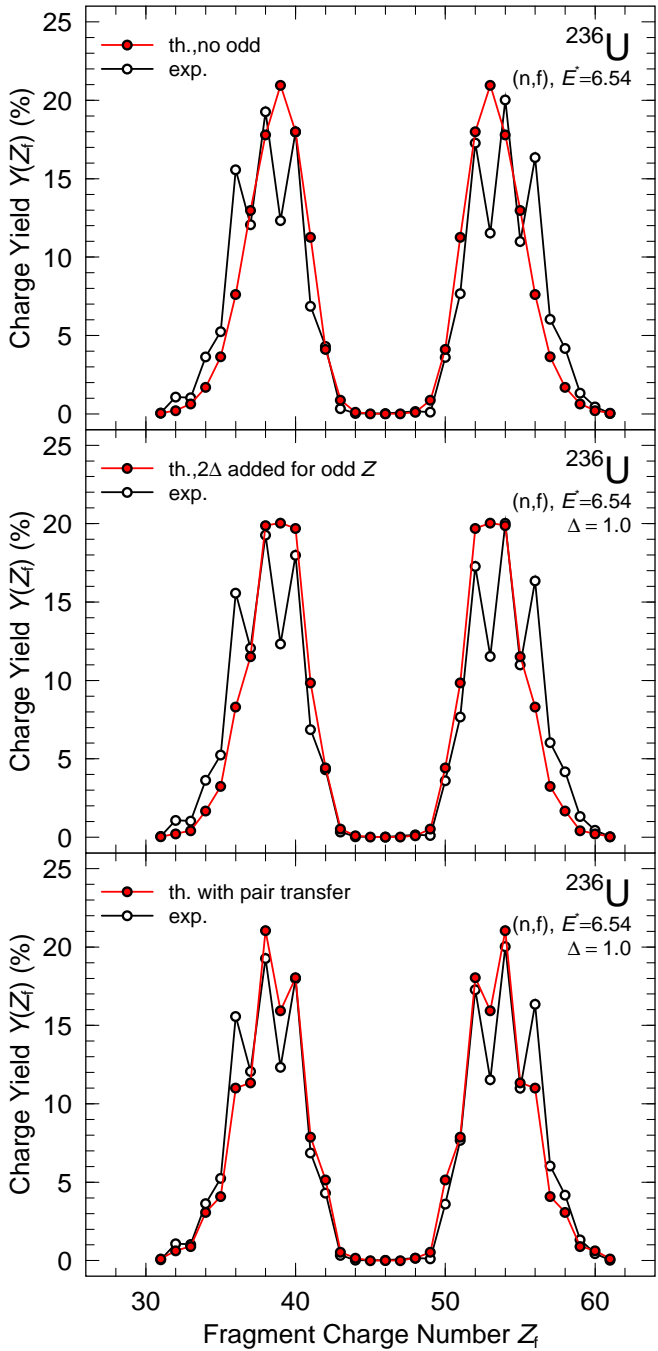}
   \caption{As Fig.~\ref{yields1} but for
$^{236}$U. \vspace{-0.4in} }
\label{yields3}
 \end{center}
\end{figure} 	
\begin{figure}[t]
 \begin{center}
\includegraphics[width=\linewidth]{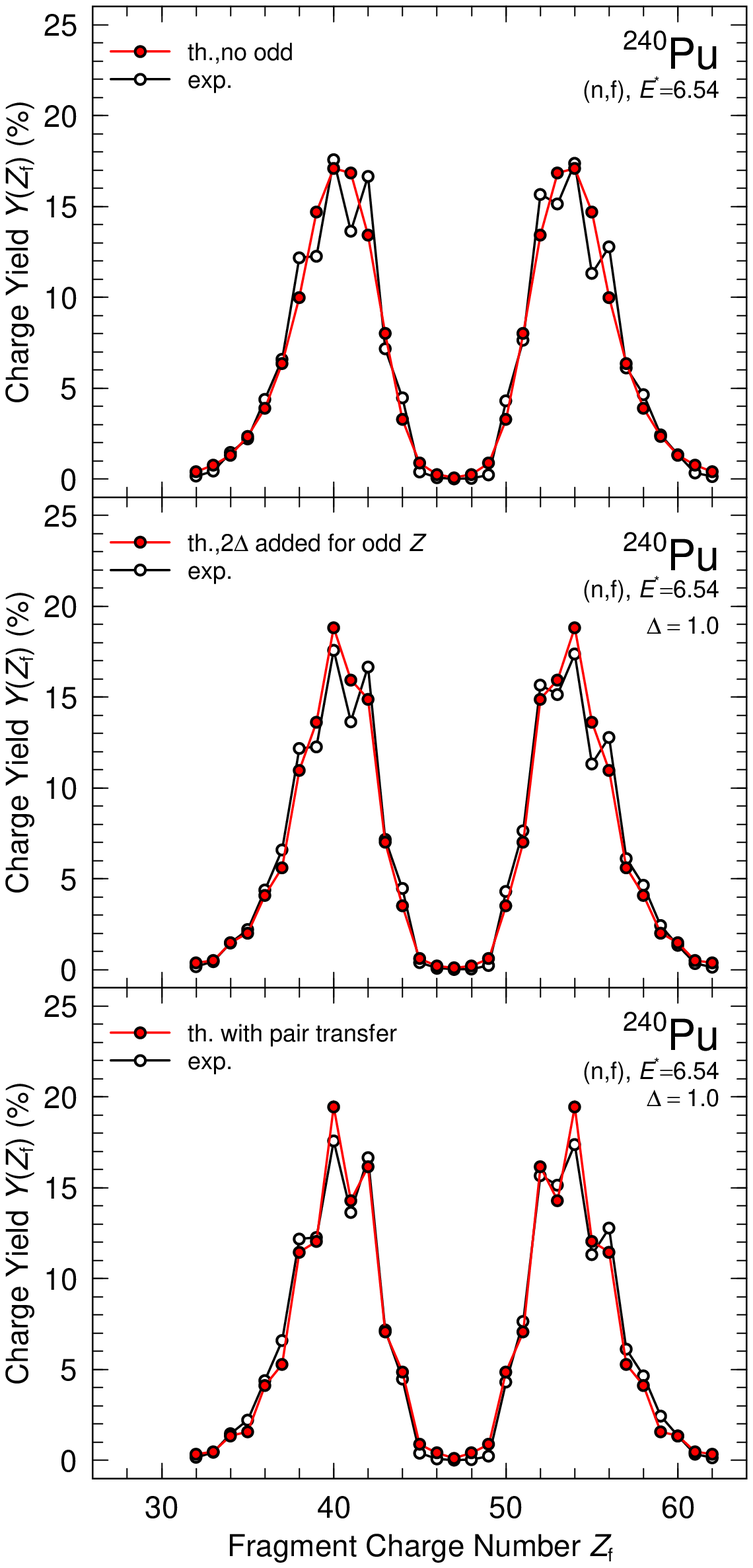}
   \caption{As Fig.~\ref{yields1} but for $^{240}$Pu. \vspace{-0.4in}}
\label{yields4}
 \end{center}
\end{figure} 	
 
\section{Application to Charge Asymmetries}
A calculation of the
complete (2D) $Y(Z,N)$ yield distribution based on the above model  would lead
to  much more complex calculations
compared to our
current 5D implementation \cite{randrup11:a,randrup11:b,randrup13:a},
 because of its 6D nature
and associated vastly increased storage requirements,
But, as a test of the above approach,
 we can study many of its aspects by
looking at  the odd-even staggering in charge distributions without
treating the neutrons separately.  We obviously then have to calculate
the potential energy for field asymmetries $\alpha_{\rm g}$ that correspond to integer
spacing of the proton number $Z_1^{\rm s}$. For $^{240}$Pu, for each combination
of the other 4 shape variables \cite{moller09:a}, we calculate the
potential energies for asymmetries corresponding to charge splits
47/47, 46/48, 45/49 \dots, and accordingly for other elements.  This
corresponds to ``averaging'' or ``summing'' over the neutron variable,
a procedure we assume has limited effects on the charge distributions
obtained relative to summing over $N$ a calculated
complete 2D $Y(Z,N)$ distribution.  Thus, we obtain in strict analogy with our previous results,
on a discreet grid the 5D potential-energy matrix
$E_{\rm pot}(I_1,I_2,I_3,I_4,I_5)$ where
$I_1$ corresponds to  the quadrupole moment (or elongation),
$I_2$ to  the neck radius,
$I_3$ to spheroidal deformations of  one of the emerging fragments,
$I_4$ to  spheroidal deformations of the other emerging fragment,
$I_5$ to  charge asymmetries $Z_1^{\rm s}/Z_2^{\rm s}$, where the charge numbers
are integers. However, since the calculated energy, $E_{\rm comp}$, does not
contain contributions from pairing effects in the emerging fragments we
add the shape-dependent odd enhancement according to Eq.\ (\ref{pair}).
It turns out very few modifications of the random walk code are
required for this calculation.
\begin{figure}[t]
 \begin{center}
\includegraphics[width=0.99\linewidth]{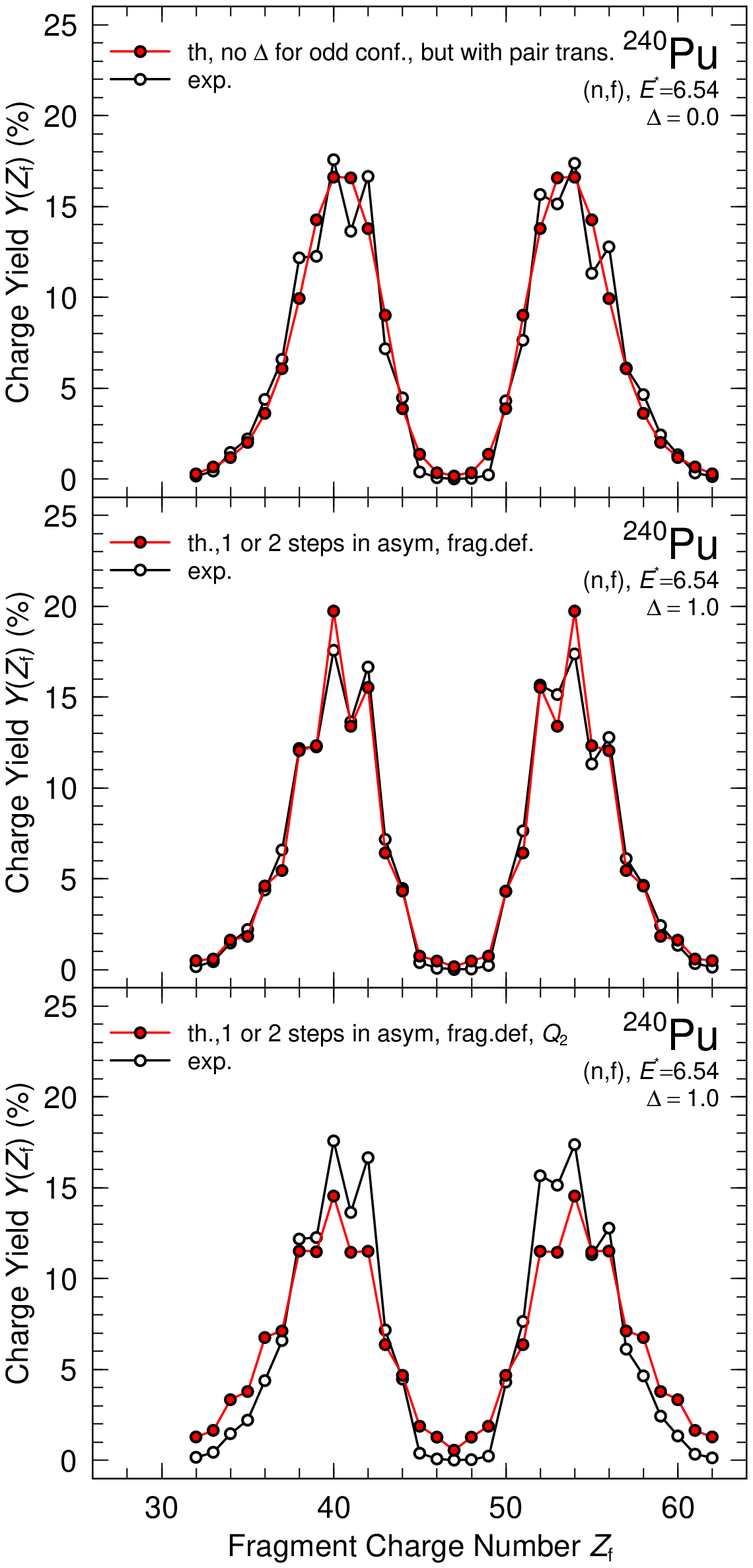}
   \caption{Calculated charge yield for neutron-induced fission of
$^{240}$Pu in the BSM model for different assumptions
about pairing and  next-step grid-point
candidates.
In the top panel no pairing effect is added to the potential energy for odd
splits, but one or two steps in asymmetry is implemented. In the middle
panel we add a pairing effect to odd splits and permit as next step candidates
one or two grid points in asymmetry and both of the fragment deformations.
In the bottom panel we have also allowed one or two steps in the elongation
coordinate $Q_2$.
\vspace{-0.2in}}
\label{yields5}
 \end{center}
\end{figure} 	
\begin{figure}[t]
 \begin{center}
\includegraphics[width=0.99\linewidth]{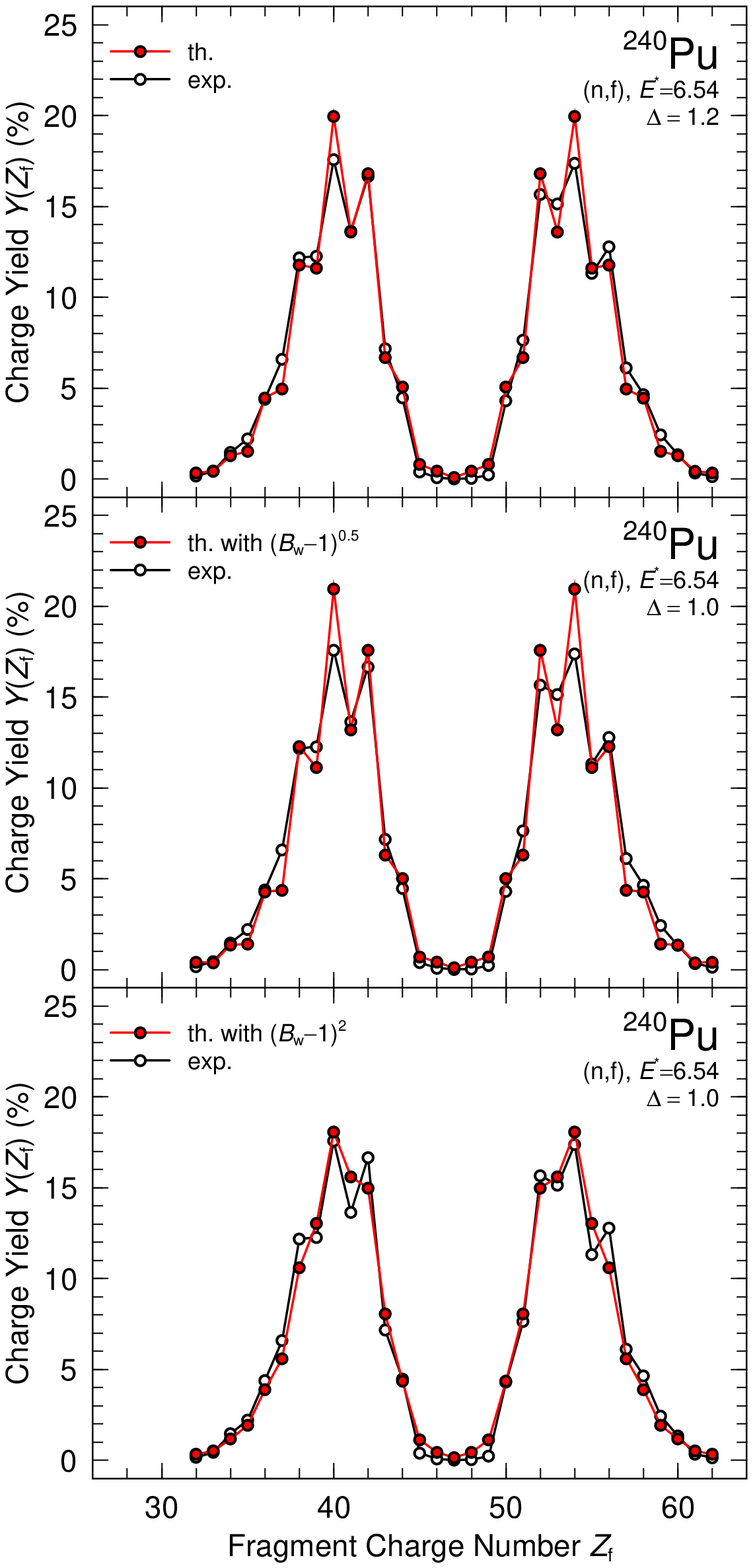}
   \caption{Study of yield sensitivity to the magnitude of the pairing $\Delta$
and the onset of fragment pairing effects. Except for the very small pairing effects
on which the bottom panel is based the results are quite robust and similar to
the standard result in the bottom panel in Fig.\ \ref{yields4}.
\vspace{-0.3in}}
\label{sensi}
 \end{center}
\end{figure} 	
 
In the BSM model we find the yield distributions by generating paths
through the potential-energy matrix as follows. We usually start the
path at the ground state (black dot in
the schematic 2D Fig.~\ref{grid}).  We then
select randomly one of the 242 surrounding points
(circles with gray
interior in Fig.~\ref{grid}, only 8 of them in this
schematic representation) as a candidate for the next point on the
track. Suppose the energy difference between this point and the
current point is $\Delta V$. Then, if $\Delta V<0$ the candidate point
becomes the next point (and current point ``at the next throw of the dice'') on the path. However, if
$\Delta V>0$ this outcome is only selected with probability $P=
\exp(-\Delta V/T)$ where $T$ is the temperature; full details are in
Refs.~\cite{randrup11:a,randrup11:b}.  When the critical neck radius
$c_0 = 2.5$ fm is reached the walk is terminated and the asymmetry of
the shape recorded.  Each yield curve is based on 20000 tracks.  We
present results for four reactions: thermal neutron-induced fission of
$^{234}$U, $^{236}$U, $^{240}$Pu, and photon-induced fission with
energies centered around 11 MeV
for $^{234}$U in Figs.\ \ref{yields1}, \ref{yields2},
\ref{yields4},  and \ref{yields3} respectively.  The experimental data
for the (n,f) reactions are from Ref.\ \cite{chadwick06:a}, the
($\gamma$,f) data from Ref.\ \cite{schmidt00:a}. The top frame in
each of the four figures is with the original model with no pairing
effect added. The middle frames are based on BSM in the 5D potential
modified according to Eq.\ (\ref{pair}).  We find little staggering in
the calculated curves, although we implemented the ``standard''
explanation for odd-even staggering: for odd-odd splits the potential
is on average $2\times \Delta$ (for zero neck radius; less in
earlier stages of the fission process) higher than the potential for even
splits. But it is  clear that the original formulation of the BSM model
would never be able to describe odd-even staggering, even after the
addition of a pairing effect to the calculated potential energy. To
illustrate why let us look at Fig.~1 and specifically at $Z=38$ on the
experimental curve. Let us assume that $Z=38$ is the asymmetry of the
current point on our evolving track.  Because we only consider
next-neighbor gridpoints as candidate points for the next point on the
path, we have to populate $Z=39$ which has a very low yield, to
subsequently populate the high-yield $Z=40$ point, for example.  To
obtain a pronounced staggering with these model features is
impossible.  But as the shape evolves in the asymmetry direction and
two levels cross it is reasonable that an alternative to breaking a
pair and transferring only one proton between the evolving fragments
is that a paired two-proton configuration could be transferred in one
step.  We have implemented this possibility in the BSM model by also
allowing $Z-2$ and $Z+2$ as next track-point candidates (shown as
circles partially filled with gray in Fig.~\ref{grid}). As alluded to above
transfer reactions indicate that correlated transfer of nucleon pairs
are common see Refs.\ \cite{broglia12:a,oertzen01:a,corradi11:a}.
In our current treatment transfer of either a
paired configuration or breaking of a pair and transfer
of one proton  can occur.  Typical excitation energies
at the end of the asymmetric tracks, see Ref.~\cite{moller14:b} are
7.7--11.6 MeV, and lower earlier in the shape evolution.  Our
consideration of paired configurations is quite consistent with the
results of Ref.\ \cite{uhrenholt13:a}, where in one example at 8.4 MeV
excitation the proportion of paired configurations is 36\%.

The calculated yields with transfers of correlated pairs allowed as
next track-point candidates are in the bottom panels in
Figs.~\ref{yields1}--\ref{yields4}. Staggering is now obvious in the
calculated results and in close agreement with the experimental data.
In the calculated results in Figs.\ \ref{yields1}--\ref{yields4} it
may seem that the crucial generalization that we introduced to
describe the odd-even staggering is not the first step
we took, namely the addition of $2\times \Delta$
to the odd charge splits, but the second step in which we permitted
a change in $Z$ of two units, corresponding to a transfer of a
paired proton configuration. But what if we had implemented this
as a first step? We have investigated this possibility and
show in the top panel of
Fig.~\ref{yields5} the calculated charge yield for neutron-induced
fission of $^{240}$Pu with no
pairing energy added to the odd charge splits, but with both one
and two steps in $Z$ permitted as next candidate points on
the random-walk trajectory. There is no  odd-even staggering in
the calculated curve which is extremely similar to the calculated yield
in the top panel in Fig.\ \ref{yields4}. It is interesting to observe
that although allowing both  one and two steps in $Z$ effectively corresponds
to increasing the speed of motion in asymmetry, or equivalently
increasing the distance between grid points, there is little change
in the calculated yield curve.
In the middle panel of Fig.\ \ref{yields5} we have
added (fractions of) 2$\Delta$ to the potential energies
for odd charge splits, allow one or two steps in asymmetry,
but also allow one or two steps in the fragment deformation
shape variables. There is little difference compared to the bottom
panel in Fig.\ \ref{yields4}.
This again shows, as was earlier pointed
out \cite{randrup11:b}
that the calculated yield curves in the BSM model do not depend
sensitively on most changes in the deformation grid.
In the bottom panel of Fig.\ \ref{yields5} we have furthermore
allowed one or two steps in the elongation variable $Q_2$.
In this case there is a noticeable effect on the calculated yield.
Now the calculated distribution is slightly wider than the experimental
results. In Ref.\ \cite{randrup11:b} we showed that tripling the number
of points in $Q_2$ led to a calculated distribution that was narrower
than the experimentally observed one. Obviously, extreme changes in the grid
will influence the calculated results. For example if we were to use
only three grid points in the elongation variable we would not obtain
realistic yields.

Finally we study, for $^{240}$Pu, the sensitivity to variations of the
postulated shape dependence that governs the
onset of odd-even effects on the potential energy
and to the magnitude of the pairing $\Delta$. The results
are shown in Fig.\ \ref{sensi}.
The top panel shows the effect of increasing the pairing
$\Delta$ by 20\% relative to our standard assumption
of $\Delta = 1.0$ MeV\@. The calculated curve is little different for the
result in the bottom panel of Fig.\ \ref{yields4}. Therefore the exact choice
of $\Delta$ is a non-issue in this first study of odd-even staggering.
In larger systematic studies a well-defined prescription should obviously
be introduced. The middle and bottom panels show the results for different
forms of the shape factor $(B_{\rm W}-1)^k$ which governs the rate with which
the effect of  the
odd-even pairing $\Delta$ in the fragments affects the calculated
potential energy as scission is approached. The quantity $B_{\rm w}$ is close to
1.5 at our selected scission configuration. Therefore, in the middle panel
the factor  $(B_{\rm W}-1)^{0.5}$ is 0.7, so that for the odd configurations
1.4 MeV is added, less earlier in the division process. The staggering here
is only very slightly larger than in our standard calculation in the bottom
panel of Fig.\ \ref{yields4} where 1.0 MeV is added in the scission region.
In the bottom panel of Fig.\ \ref{sensi} the shape factor  $(B_{\rm W}-1)^2$
comes out to 0.25 so that only 0.5 MeV or less is added to the odd-odd divisions.
Here the staggering is much reduced, as might be expected. However from these
sensitivity studies we conclude that the results are quite robust for
reasonable variations of  assumptions about the onset of fragment character
on the potential energy as well as to the magnitude of the pairing $\Delta$.
 
In summary we
have shown that to describe odd-even staggering in the
BSM model we must add simultaneously two effects: (1)
odd-even effects on the calculated potential-energy surface
and (2) allow transfers of correlated paired proton
configurations; they work together in the development of
odd-even staggering.
 
As future ``perspectives for the next decade'' we
anticipate that to develop more
accurate descriptions we need to
\begin{itemize}
\item implement the extensions we discuss here into a computer model
framework and calculate the full $Y(Z,N)$ fission-fragment yield distributions
(a two-dimensional function of neutron and proton number),
\item treat from more basic principles how the number of paired configurations
decrease with excitation energy (see  Ref.\ \cite{uhrenholt13:a}
for a discussion) which should influence the probability with which
a point corresponding to two-nucleon transfer is chosen as the next candidate point
on the trajectory,
\item calculate the damping of shell effects based on
actual single-particle structure rather than use a parameterized approach.
\item
understand issues related to the deformation grid.
Clearly the calculated yields do depend on the selection of the grid.
To take an obvious example, were we to have only 3 grid points in the elongation
direction we would not obtain any realistic yields. However, in the
extensive sensitivity studies in
Ref.\ \cite{randrup11:b} it was shown that the yields were remarkably insensitive
to many types of grid changes, as is also shown by the results of Figs.
\ref{yields5} and \ref{sensi}
above.
\end{itemize}

Discussions with A. Sierk, A. Iwamoto, and  J. Randrup are appreciated.  This work was
supported by travel grants for P.M.\ to JUSTIPEN (Japan-U.S. Theory
Institute for Physics with Exotic Nuclei) under grant number
DE-FG02-06ER41407 (U. Tennessee).  This work was carried out under the
auspices of the NNSA of the U.S. Department of Energy at Los Alamos
National Laboratory under Contract No.\ DE-AC52-06NA25396.  TI was
supported in part by MEXT SPIRE and JICFuS and JSPS KAKENHI Grant
no. 25287065.

%


\end{document}